\documentclass [12pt]{article}
\usepackage{graphics}
\usepackage{epsfig}
\usepackage{hangcaption}
\usepackage{array}
\title{Interactions of a Single Goldstino}

\author{Taekoon Lee\footnote{email: tlee@physics.purdue.edu}$^{*(a,b)}$ and
Guo-Hong Wu\footnote{email: wu@physics.purdue.edu}$^{\dagger(a)}$
         \\
  {\it Department of Physics, Purdue University, West Lafayette, IN 47907
  }$^{(a)}$\\
  {\it Department of Physics, Seoul National University,
  Seoul, 151-742, Korea}$^{(b)}$\footnote{Present address.}
  }

\date{}
\begin{document}
\maketitle
\begin{abstract}
The single goldstino interaction is given by
the goldstino derivative coupling to the supercurrent. In an alternative
description, the goldstino couples nonderivatively.
We give a simple method to establish the equivalence of the two approaches,
valid to all orders in perturbation theory, and for any scattering
process involving an arbitrary number of particles, but with a single
external goldstino. In the meantime, we find in the nonderivative 
form of the goldstino interaction a new quartic vertex that has been 
overlooked, and terms that were included incorrectly in the literature.
The phenomenological implication of this new quartic operator is 
discussed. 

\end{abstract}

\def\thepage{PURD-TH-98-09, SNUTP-98-131}
\thispagestyle{myheadings}
\newpage
\pagenumbering{arabic}
\addtocounter{page}{0}
\newcommand{\be}{\begin{equation}}
\newcommand{\ee}{\end{equation}}
\newcommand{\bear}{\begin{eqnarray}}
\newcommand{\eear}{\end{eqnarray}}

Light gravitino is common in some models of supersymmetric extension
of the standard model, as in gauge mediated SUSY breaking
models \cite{gm} and no scale supergravity models \cite{noscale}.
Phenomenologically a light gravitino is interesting because it
could be produced in systems with relatively low energies compared
to the SUSY breaking scale. An observation of a gravitino emitting process
in an accelerator experiment, for example,  could determine a very
important parameter, the SUSY breaking scale. It is thus important
to understand the gravitino interaction with other fields, for example,
those in the minimal supersymmetric standard model (MSSM).

The interaction of the helicity $\frac{1}{2}$ longitudinal component
of the gravitino with matter becomes
stronger as the gravitino mass gets smaller.
In the small mass limit, which is valid if the energy in consideration is much
bigger than the gravitino mass, one can replace using the
SUSY version of the equivalence theorem  \cite{fayet,equiv} the gravitino with
the goldstino which was eaten by it. Then because the goldstino
is the Goldstone fermion of spontaneous SUSY breaking, its
coupling with other fields is determined by the well-known
derivative coupling of goldstino to the  supercurrent, much like the
derivative coupling of pions in spontaneous chiral symmetry breaking.

On the other hand, if one  works out in a  given linearly realized
SUSY model, one usually gets goldstino interactions
in nonderivative form. In this formalism, the triple vertices
of goldstino-boson-fermion are fixed by the goldstino Goldberg-Treiman
relation  \cite{fayet,clark} in which
the couplings are proportional to the mass
 splittings of the boson-fermion pairs.

The two different forms of goldstino coupling are {\it expected} to give
identical amplitude in scatterings with a single external
goldstino because the derivative coupling is part of the
nonlinearly realized SUSY effective lagrangian \cite{ivanov,wess,clark,cllw},
which can, in principle,
be obtained from the corresponding linearly realized SUSY model by
field redefinition \cite{ivanov}. 
In practice, however, finding the field transformations can be quite 
involved \cite{luty}, and to the best of our knowledge, there is no
explicit proof of the equivalence of the two formulations of the goldstino
interaction. In this
letter, we give a simple method for establishing the 
equivalence of the derivative and nonderivative couplings of the goldstino, 
without using any field redefinitions. Our method applies to 
any scattering process with an arbitrary number of external
particles but with one external goldstino, and to all orders in
perturbation.
By using this method, one can
identify the complete set of operators for the nonderivative goldstino 
couplings in any given model.      
As a result of this exercise, we find a quartic vertex that has not
been discussed previously,
and terms that were included incorrectly in the nonderivative
formalism.

For simplicity, we apply our method to SUSY QED. However,
the method can be straightforwardly used for
more complex models, and  later we will comment
 on the goldstino interaction with the MSSM fields.
The SUSY QED we consider is comprised of a photon $A_{\mu}$,
a photino $\lambda$ with mass $m_{\lambda}$, a complex scalar
$\phi$ with mass $m_{\phi}$, and a massless Weyl fermion $\psi$.
The interaction lagrangian is given by:
\bear
{\cal L}_{\mbox{{\tiny int}}} &=& -e A_{\mu} \psi \sigma^{\mu} \overline{\psi}
+ i e A_{\mu} ( \phi^{*} \partial^{\mu} \phi -
\partial^{\mu} \phi^{*} \phi) \nonumber \\
                      && - \sqrt{2} e (  \phi^{*}\psi\lambda +
\phi\bar{\psi}\bar{\lambda}) -\frac{e^{2}}{2} (\phi^{*}\phi)^
{2} - e^{2} A_{\mu}A^{\mu}\phi^{*}\phi
\label{e1}
\eear
where $e$ is the $U(1)$ gauge coupling and $\psi$ and $\phi$
carry a unit charge.
Throughout this paper we follow
the convention for spinors and metric given in \cite{bagger}, except
that our gaugino $\lambda$ is related to the gaugino
$\lambda_{WB}$ of \cite{bagger} by $\lambda= -i \lambda_{WB}$.
Though we have chosen, for simplicity, a model with only one Weyl spinor,
which is not anomaly free, our method applies to Dirac
fermions as well. We will come back to this later.

The derivative coupling of the goldstino to the supercurrent is
given by
\be
{\cal L}_{\mbox{{\scriptsize D}}}= \frac{1}{F} \partial_{\mu}\chi^{\alpha}
J_{\alpha}^{\mu} + h.c.
\label{e2}
\ee
where $\chi$ denotes the goldstino, $F$ is the goldstino decay constant,
and $J^{\mu}$ is the supercurrent,
\be
J^{\mu}= \sigma^{\nu}\bar{\sigma}^{\mu}\psi D_{\nu}\phi^{*}
-\frac{1}{2\sqrt{2}}\sigma^{\nu}\bar{\sigma}^{\rho}\sigma^{\mu}
\overline{\lambda} F_{\nu\rho} \;\; ,
\label{e3}
\ee
with
\be
D_{\mu}\phi= (\partial_{\mu}+i e A_{\mu} ) \phi, \hspace{.5 cm}
F_{\mu\nu}=\partial_{\mu}A_{\nu}-\partial_{\nu}A_{\mu}.
\ee

In the nonderivative form of the goldstino interaction, though the triple
vertices are fixed by the Goldberg-Treiman relation, the quartic
vertex (or vertices) has not been thoroughly explored.
However, our proof of the equivalence of the derivative and nonderivative
descriptions of the single goldstino interaction gives
the following nonderivative lagrangian,
\be
{\cal L}_{\mbox{{\scriptsize ND}}} = \frac{m_{\phi}^{2}}{F} \chi \psi\phi^{*} +
\frac{im_{\lambda}}{\sqrt{2}F} \chi \sigma^{\mu\nu}\lambda F_{\mu\nu}
-\frac{ e m_{\lambda}}{\sqrt{2}F} \phi^{*}\phi\chi\lambda +h.c. \;\; ,
\label{e5}
\ee
where the first two are the standard terms, while the quartic term
will be justified later.
We now  prove that the lagrangians  (\ref{e2}) and (\ref{e5}) give identical
amplitude to an arbitrary order in gauge coupling for any scattering process
involving a single external goldstino.
For this purpose, let us consider the difference between the two lagrangians,
\bear
\delta{\cal L} &=& {\cal L}_{\mbox{{\scriptsize D}}}-
{\cal L}_{\mbox{{\scriptsize ND}}} \nonumber \\
&=&\overline{{\cal L}}_{1}+ \delta V_{4} + {\cal L}_{\mbox{{\scriptsize
gauge}}} + ( \mbox{total derivative}) \;\; ,
\eear
where
\bear
\overline{{\cal L}}_{1} &=& \frac{1}{F}(\partial^{2} -
m_{\phi}^{2})\phi^{*}\chi\psi -\frac{1}{F} \partial^{2}\psi\chi\phi^{*}
\nonumber \\
&&-\frac{1}{2 \sqrt{2}F}( \partial_{\mu}\bar{\lambda} \bar{\sigma}^{\mu}
 +im_{\lambda}{\lambda}) \sigma^{\rho}\bar{\sigma}^{\nu} \chi F_{\nu\rho}
\nonumber \\
&& +\frac{1}{\sqrt{2}F} (\partial^{2}g_{\mu\nu}- 
(1-\frac{1}{\xi}) \partial_{\mu} \partial_{\nu})
A^{\nu} \chi\sigma^{\mu}\overline{
\lambda} \;\; + h.c.
\label{e8}
\eear
and
\be
\delta V_{4} =-\frac{i e}{F} \partial_{\mu} \chi \sigma^{\nu}
\bar{\sigma}^{\mu} \psi \phi^{*} A_{\nu} +
\frac{ e m_{\lambda}}{\sqrt{2}F} \phi^{*}\phi\chi\lambda
 \;\; + h.c. \;\; ,
\ee
\be
{\cal L}_{\mbox{{\scriptsize gauge}}} =
-\frac{1}{\sqrt{2}\xi F}\chi\sigma^{\mu}\bar{\lambda}\partial_{\mu}
\partial_{\nu}A^{\nu} + \;\;h.c..
\ee
Here $\xi$  is  the  gauge fixing parameter, and the corresponding  photon
propergator is given by
\be
D_{\mu\nu}(q)= -\frac{i}{q^{2}}\left(g_{\mu\nu} -(1-\xi)\frac{q_{\mu}
q_{\nu}}{q^{2}} \right)\;\; .
\ee
Note that each term in $\overline{{\cal L}}_{1}$ is proportional 
to the free field
equation, and thus vanishes on-shell,
and ${\cal L}_{\mbox{\scriptsize gauge}}$ is proportional to the 
divergence $\partial_{\mu}A^{\mu}$.

To prove the equivalence, we need to show the following matrix element
between arbitrary initial and final states vanishes,
\bear
\delta S_{fi} &=& <f|\mbox{{\bf T}} e^{ i\int dx {\cal L}_
{\mbox{{\tiny int}}}(x)}
\int dx \delta {\cal L}(x) |i> \nonumber \\
&=&<f|\mbox{{\bf T}} e^{ i\int dx {\cal L}_
{\mbox{{\tiny int}}}(x)}\left[\int dx
\overline{{\cal L}}_{1}(x)+ \int dx \delta V_{4}(x)\right]|i> \nonumber \\
&=& 0 \; ,
\label{e9}
\eear
where {\bf T} denotes the time-ordered product. Note that here we 
ignored ${\cal L}_{\mbox{\scriptsize gauge}}$ term because its contribution
to $\delta S_{fi}$ vanishes due to the Ward identity. 
Now because it vanishes when a field in $\overline{{\cal L}}_{1}$, proportional
to the
free field equation, contracts with an external particle, nonvanishing term
can arise only when the field contracts to become an internal line in a Feynman
diagram. And because of the free field equation, when
such a field contracts, it generates a local operator.
This implies that
the difference in amplitude between the two formalism is given by
local operator insertions. We  will show that the sum of such local
operator insertions
generated by
$\overline{{\cal L}}_{1}$ vanishes when combined with $\delta V_{4}$.

Consider
\bear
&&<f|\mbox{{\bf T}} e^{ i\int dx {\cal L}_{\mbox{{\tiny int}}}\left(x\right)}
\int dx' \overline{{\cal L}}_{1}(x')|i> \nonumber \\
&=&<f|\mbox{{\bf T}} \sum_{n=0}^{\infty} \frac{ \left[i
\int dx {\cal L}_{\mbox{{\tiny int}}}
(x)\right]^{n}}{n!}\int dx' \overline{{\cal L}}_{1}(x')|i> \nonumber \\
&=&<f|\mbox{{\bf T}} \sum_{n=1}^{\infty} \frac{ [i \int dx {\cal L}_
{\mbox{{\tiny int}}}(x)]^
{n-1}}{(n-1)!}\int dxdx' \ll i
{\cal L}_{\mbox{{\tiny int}}}(x),\overline{{\cal L}}
_{1}(x')\gg|i>  \nonumber \\
&=&<f|\mbox{{\bf T}} e^{ i\int dx {\cal L}_{\mbox{{\tiny int}}}\left(x\right)}
\int dxdx' \ll i {\cal L}_{\mbox{{\tiny int}}}(x),
\overline{{\cal L}}_{1}(x')\gg|i> \;\; ,
\label{e10}
\eear
where $\ll{\cal O}_{1}(x),{\cal O}_{2}(x')\gg$ is a local operator
obtained by contracting the field in ${\cal O}_{2}$ that is proportional to
the free field equation with the corresponding one in  ${\cal O}_{1}$.
For example, if
\be
 {\cal O}_{1}(x)= \bar{\lambda}\bar{\psi}\phi(x)  
 \;\;\;\;\;\;\;\;\;\;\;\;\;\;\;\;\;
 {\cal O}_{2}(x')=(\partial^{2}-m_{\phi}^{2})\phi^{*} \chi\psi(x') \; ,
\ee
then
\bear
\ll{\cal O}_{1}(x),{\cal O}_{2}(x')\gg&=& \bar{\lambda}\bar{\psi}(x)<\phi(x)
(\partial^{2}-m_{\phi}^{2})\phi^{*}(x')> \chi\psi(x') \nonumber \\
&=& i \bar{\lambda}\bar{\psi}\psi\chi(x)\delta(x-x')\,\,.
\eear
\begin{figure} 
\begin{center}
\epsfig{file=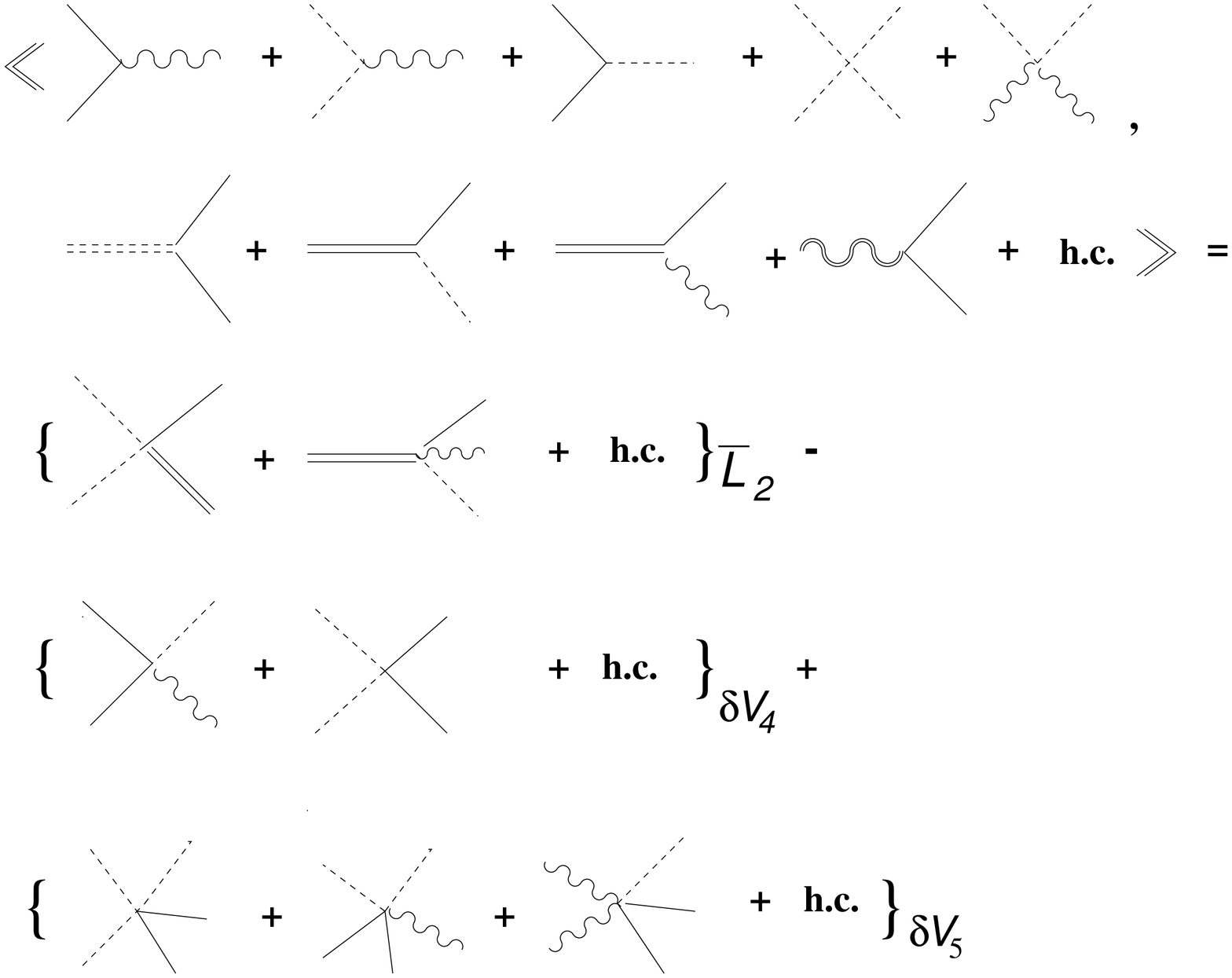, %
        height=10.0cm, angle=-90}
\end{center}
\isucaption{Local operators generated by the contraction 
$ \ll ~i {\cal L}_{\mbox{{\tiny int}}}(x), \overline{{\cal L}}_{1}(x')\gg$
(see Eq.~(\ref{e13})). The solid, dashed, and wavy lines denote fermions,
scalars, and gauge bosons respectively.  Double lines denote fields 
proportional to their respective free field equations. }
\end{figure}
Using (\ref{e1}) and (\ref{e8}), we get (see Fig.1)
\be
\int dxdx' \ll i {\cal L}_{\mbox{{\tiny int}}}(x),
   \overline{{\cal L}}_{1}(x')\gg=
\int dx (\overline{{\cal L}}_{2}(x)-\delta V_{4}(x) +\delta V_{5}(x)) \;\; ,
\label{e13}
\ee
where
\be
\delta V_{5}= \frac{e^{2}}{F} \phi^{*}\phi \phi^{*}\psi\chi
 + \frac{\sqrt{2}e^2}{F} A_{\mu} \phi^{*}\phi \chi\sigma^{\mu}\bar{\lambda}
+ \frac{e^2}{F} A_{\mu}A^{\mu} \phi^{*} \psi \chi +\;\; h.c. \;\; ,
\ee
and
\be
\overline{{\cal L}}_{2}= \frac{ie}{\sqrt{2}F} \phi^{*}\phi
 \chi (\sigma^{\mu} \partial_{\mu}\bar{\lambda}- i m_{\lambda}\lambda)
+\frac{i e}{F} \partial_{\nu}\psi\sigma^{\nu}\bar{\sigma}^{\mu}
\chi\phi^{*} A_{\mu} +\;\; h.c. \;\; .
\ee
As with $\overline{{\cal L}}_{1}$, every term in $\overline{{\cal L}}_{2}$
is proportional to a  free field equation, and
thus vanishes on-shell.
Using (\ref{e9}),(\ref{e10}) and (\ref{e13}) ,
\be
\delta S_{fi}=<f|\mbox{{\bf T}} e^{ i\int\,dx {\cal L}_
{\mbox{{\tiny int}}}}\left[
\int dx \overline{{\cal L}}_{2}(x)+ \int dx \delta V_{5}(x)\right]|i>\,\,.
\ee

Now repeating the same manipulation with $\overline{{\cal L}}_{2}$ we have
(see Fig.2)
\bear
\delta S_{fi}&=&<f|\mbox{{\bf T}}e^{ i\int\,dx {\cal L}_
{\mbox{{\tiny int}}}}\left[
\int dxdx' \ll i {\cal L}_{\mbox{{\tiny int}}}(x),
\overline{{\cal L}}_{2}(x')\gg
+ \int dx \delta V_{5}(x)\right]|i> \nonumber \\
&=&0 \;\; ,
\eear
because
\be
\ll i {\cal L}_{\mbox{{\tiny int}}}(x),\overline{{\cal L}}_{2}(x')\gg
=-\delta V_{5}(x)
\delta(x-x')\,\,.
\label{e18}
\ee
This completes the proof. We have thus shown that the difference
in amplitude is either given by contractions of a field that is
proportional to its free field equation with an external particle, or
 insertions of null operators, or both. In any case,
it vanishes.  Note that the proof is not only
true at tree level, but also in any order of loop expansion
because the steps we have taken through (\ref{e9})--(\ref{e18})
are applicable with
a given regularization of loop divergences.

\begin{figure} 
\begin{center}
\epsfig{file=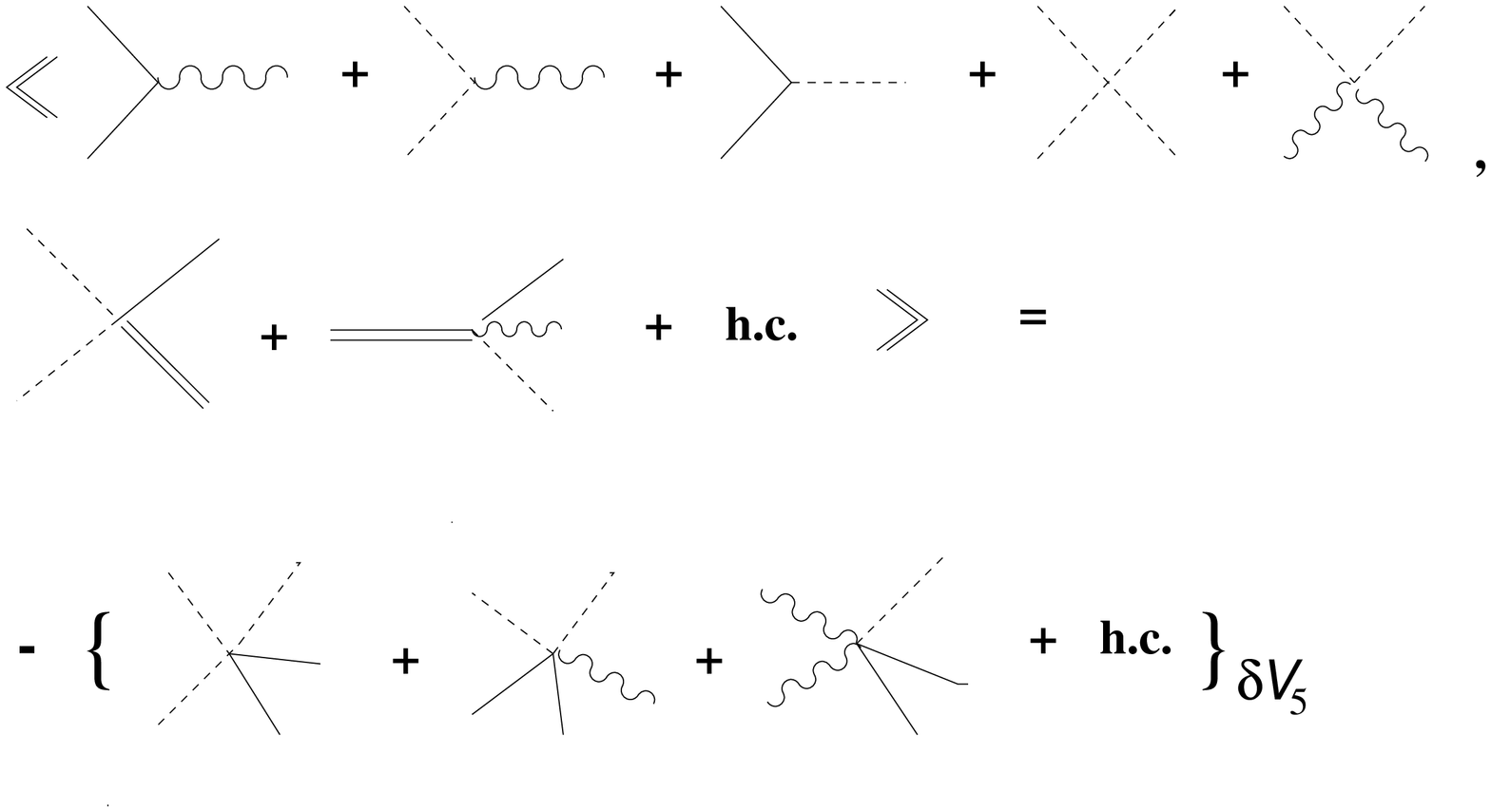, %
        height=10.0cm, angle=-90}
\end{center}
\isucaption{ Local operators generated by the contraction 
$ \ll i {\cal L}_{\mbox{{\tiny int}}}(x),
   \overline{{\cal L}}_{2}(x')\gg$ (see Eq.~(\ref{e18})). 
The solid, dashed, and wavy lines denote fermions,
scalars, and gauge bosons respectively.  Double lines denote fields 
proportional to their respective free field equations. }
\end{figure}

Thus far we have only considered a model with a Weyl fermion.
For SUSY QED with a Dirac fermion with mass $m_{\psi}$,
the derivative coupling of the goldstino to the supercurrent can be
easily obtained from (\ref{e2}) and (\ref{e3}) with  $\psi \rightarrow \psi_i$
and $\phi \rightarrow \phi_i$, where $i=+,-$ denotes the charges of the
fields.
Using the above method, the non-derivative coupling of the goldstino
is found to be,
\bear \label{eq:QED}
{\cal L}_{\mbox{{\scriptsize ND}}} & = &
 \frac{m^2_{\phi_i} - m^2_{\psi}}{F} \chi \psi_i \phi_i^*
 + \frac{im_{\lambda}}{\sqrt{2}F} \chi \sigma^{\mu\nu} \lambda F_{\mu\nu}
 - \frac{em_{\lambda}}{\sqrt{2}F} \chi \lambda (\phi_+^* \phi_+
                                               -\phi_-^* \phi_-)
     \;\; + h.c. \;\; ,
\eear
where summation over $i$ is implied and $\phi_+$ and $\phi_-$ are
taken to be mass eigenstates.

 The above analysis can be straightforwardly generalized to 
nonabelian gauge theories.  The nonderivative goldstino coupling
 for SUSY QCD with one quark flavor is given by, 
\bear \label{eq:QCD}
{\cal L}^{QCD}_{\mbox{{\scriptsize ND}}} & = &
 \frac{m^2_{\phi} - m^2_{\psi}}{F} \chi \psi \phi^*
 + \frac{im_{\lambda}}{\sqrt{2}F} \chi \sigma^{\mu\nu} \lambda^a 
F^a_{\mu\nu}
-\frac{g m_{\lambda}}{\sqrt{2}F} \chi \lambda^{a}
\phi_{i}^{*}T^{a}_{ij}
\phi_{j} \; + h.c. \;\; ,
\eear
where $g$ is the gauge coupling and  $T^{a}$ are the  generators of 
the gauge group. Note that the presence of the third term is again
required.

Now a comment is in order on the nonderivative
form of the goldstino interactions (\ref{e5}), (\ref{eq:QED}), 
and (\ref{eq:QCD}).
Without the quartic vertex, the equivalence would have failed, and so
it must be included in the nonderivative form. The nonabelian version
\be \label{eq:quartic}
-\frac{g m_{\lambda}}{\sqrt{2}F} \chi \lambda^{a}
\phi_{i}^{*}T^{a}_{ij}
\phi_{j}\,\,,
\ee
should be present in the nonderivative form of
goldstino interaction with the MSSM fields, but was missing
in Ref. \cite{nano}, and was overlooked in Ref. \cite{moroi}. 
Since this term must exist model-independently, it is  convincing to know 
that the model discussed in Ref. \cite{zwirner} does
have the quartic vertex with the  right 
coefficient. Also note
the absence of $A_{\mu}\chi\psi\phi^{*}$ vertex in the nonderivative form,
while it exists in the derivative form. Terms of this type in the
derivative form in MSSM were mistakenly included in the nonderivative
goldstino coupling in Ref. \cite{nano} ( (e),(f),(e'), and (f') of Table 1  in
the reference).
We also note that existence of this term in
the nonderivative form of goldstino interaction
would violate the gauge symmetry, and thus should not be allowed.
The goldstino phenomenology studied in  Ref.~\cite{nano} is, however,
 not dependent on the quartic term  (\ref{eq:quartic}).

  The operator (\ref{eq:quartic}) contributes directly
to the goldstino emission rate through 
$ \phi_i + \phi_j^* \rightarrow \chi + \lambda^a$ and 
$\phi_i + \lambda^a \rightarrow \chi + \phi_j$, where
$ \phi_i$ and $\lambda^a$ stand for a squark and a gluino respectively.
These two are part of the goldstino production processes 
considered in Ref.~\cite{moroi} to derive the constraint on the 
light gravitino mass from cosmology.
To see the effect of this quartic operator, we have computed
the cross sections for these two processes in the limit 
when the center of mass energy is much bigger than the squark and 
gluino masses, as considered in Ref.~\cite{moroi}.
In this limit, only the dimension-5 operators in (\ref{eq:QCD}) need be
considered.
 
  For $ \phi_i + \phi_j^* \rightarrow \chi + \lambda^a$, we find that 
the $s$-channel gluon exchange contribution to the cross section 
is $(g^2m^2_{\lambda}/48\pi F^2)T^a_{ji} T^{a\,*}_{ji}$, 
in agreement with the result
of Ref.~\cite{moroi}. The quartic term contributes 
$(g^2m^2_{\lambda}/16\pi F^2)T^a_{ji} T^{a\,*}_{ji}$, 
which is three times as large.
The interference term vanishes identically, and the total cross
section is simply given by
\bear
\sigma( \phi_i + \phi_j^* \rightarrow \chi + \lambda^a) & = &
\frac{g^2m^2_{\lambda}}{12\pi F^2}T^a_{ji} T^{a\,*}_{ji} \;,
\eear 
which is 4 times of that given in  Ref.~\cite{moroi}.

  For the process $\phi_i + \lambda^a \rightarrow \chi + \phi_j$,
the $t$-channel gluon exchange contribution to the cross section 
depends on the cut imposed on the scattering angle, as obtained
in  Ref.~\cite{moroi}. We find that the quartic term (\ref{eq:quartic}) 
contribution is given by 
$(g^2m^2_{\lambda}/64\pi F^2)T^a_{ji} T^{a\,*}_{ji}$,
which is small compared to the gluon exchange contribution for the angular cut
used in Ref.~\cite{moroi}.

The effect of the quartic operator on the total cross section 
for goldstino production after summing over all processes is expected
to be at the few per cent level, depending on the angular cut used.
The correction to the cosmological bound on the light gravitino mass 
is correspondingly small.

  Before we conclude, we would like to remark that although 
both the derivative and the nonderivative forms of goldstino coupling
have been used for over two decades, an explicit proof of the equivalence 
of the two formulations has not been available to the best of our knowledge. 
In fact, it seems that some discrepancies exist regarding the complete set 
of operators in the nonderivative form\cite{moroi,nano}.
In this letter, we have given the complete set of operators associated with
the single goldstino nonderivative coupling in SUSY QED and QCD. 
In particular, we point out one quartic operator that has been 
missing in many discussions of goldstino phenomenology.
More generally, it is important to have a simple and straightforward 
method available to identify the complete set of operators for the 
nonderivative goldstino interaction in
any given model, and we hope that our method fits into this category.
\\\

We thank T. Clark and S. Love for useful conversations. This work was
supported in part by the U.S. Department of Energy under grant
DE-FG02-91ER40681 (task B). Also one of the authors (T. L.) is  
supported in part by the KOSEF brain pool program.

 \end{document}